\documentclass[11pt]{article}
\usepackage[a4paper,margin=0.85in]{geometry}
\usepackage{amsmath}
\usepackage{booktabs}
\usepackage{graphicx}
\usepackage{url}
\usepackage{xurl}
\usepackage{array}
\usepackage{caption}
\usepackage[super,sort&compress]{natbib}
\usepackage{setspace}
\usepackage{lineno}
\newif\ifsubmission
\ifdefined\journalsubmission\submissiontrue\fi
\ifsubmission
\linenumbers
\fi

\newcommand{\compPctLow}{73}
\newcommand{\compPctHigh}{92}
\newcommand{\compPanels}{18}

\newcommand{\forcedOneIn}{31}

\newcommand{\forcedSeasons}{27}

\newcommand{\absenceInjuryPct}{39}
\newcommand{\absenceTacticalPct}{4}
\newcommand{\injuryPlayers}{5942}

\newcommand{\injuryAppearances}{296,029}

\newcommand{\gapShortFrom}{8}
\newcommand{\gapShortTo}{13}

\newcommand{\ofGapShortLoaded}{0.88}
\newcommand{\ofGapShortLoadedLow}{0.82}
\newcommand{\ofGapShortLoadedHigh}{0.94}

\newcommand{\ofKeeperStarts}{19,231}

\newcommand{\otherSportsRead}{seven}
\newcommand{\allSports}{nine}
\newcommand{\otherLeagues}{37}

\newcommand{\otherWithinBelow}{54}

\newcommand{\shareGroups}{11}
\newcommand{\shareSports}{six}

\newcommand{\bleagueInjuries}{27}

\newcommand{\headAllMedianMrr}{1.83}

\newcommand{\headWomen}{748}

\newcommand{\ofGapShortNarrow}{0.94}
\newcommand{\ofGapShortNarrowLow}{0.85}
\newcommand{\ofGapShortNarrowHigh}{1.05}

\newcommand{\gapTwoThreeNarrow}{1.40}
\newcommand{\gapTwoThreeNarrowLow}{1.23}
\newcommand{\gapTwoThreeNarrowHigh}{1.59}

\newcommand{\playersDistinct}{22,573}

\newcommand{\ofGapShortLoadedLowerPct}{12}

\newcommand{\nullAdvisedLow}{1.06}
\newcommand{\nullAdvisedHigh}{1.09}

\newcommand{\nullAcrossSelection}{0.92}

\newcommand{\nullAcrossLow}{0.97}
\newcommand{\nullAcrossHigh}{0.99}

\newcommand{\gapWithin}{1.06}
\newcommand{\gapWithinLow}{0.99}
\newcommand{\gapWithinHigh}{1.13}

\newcommand{\gapWithinNarrow}{1.16}
\newcommand{\gapWithinNarrowLow}{1.03}
\newcommand{\gapWithinNarrowHigh}{1.29}

\newcommand{\absLoadedStarterPct}{41}

\newcommand{\absLongGapPct}{38}

\newcommand{\gapCompletedWithin}{0.99}
\newcommand{\gapCompletedWithinLow}{0.91}
\newcommand{\gapCompletedWithinHigh}{1.09}
\newcommand{\gapCompletedAcross}{1.02}
\newcommand{\gapCompletedAcrossLow}{0.93}
\newcommand{\gapCompletedAcrossHigh}{1.11}

\newcommand{\gapCompletedWithinNarrow}{0.95}
\newcommand{\gapCompletedWithinNarrowLow}{0.83}
\newcommand{\gapCompletedWithinNarrowHigh}{1.10}
\newcommand{\gapCompletedAcrossNarrow}{0.96}
\newcommand{\gapCompletedAcrossNarrowLow}{0.84}
\newcommand{\gapCompletedAcrossNarrowHigh}{1.10}
\newcommand{\nullCompletedLow}{0.99}
\newcommand{\nullCompletedHigh}{1.01}

\newcommand{\nullCompletedAcrossLow}{1.00}
\newcommand{\nullCompletedAcrossHigh}{1.01}

\newcommand{\heavyCompletedPct}{95}
\newcommand{\longGapCompletedPct}{48}

\newcommand{\nullTwoThreeNarrowLow}{1.04}

\newcommand{\nullWithinFootballLow}{1.06}
\newcommand{\nullWithinFootballHigh}{1.17}

\newcommand{\shortCompletedNarrow}{1.41}
\newcommand{\shortCompletedNarrowLow}{1.14}
\newcommand{\shortCompletedNarrowHigh}{1.73}

\newcommand{\nullShortCompletedNarrowLow}{0.99}
\newcommand{\nullShortCompletedNarrowHigh}{1.01}

\newcommand{\knockMaxPct}{12}

\newcommand{\heavyWeeklyPct}{64}

\newcommand{\nullSeasonAllLow}{1.05}
\newcommand{\nullSeasonAllHigh}{2.19}
\newcommand{\nullSeasonOtherLow}{1.05}
\newcommand{\nullSeasonOtherHigh}{2.19}

\newcommand{\shortCompletedExcessPerThousand}{4.2}

\newcommand{\shortCompletedExcessClubSeasons}{10.5}
\newcommand{\shortCompletedExcessPerThousandLow}{1.8}
\newcommand{\shortCompletedExcessPerThousandHigh}{6.1}

\newcommand{\teamFootball}{1.19}
\newcommand{\teamFootballLow}{1.03}
\newcommand{\teamFootballHigh}{1.36}

\newcommand{\nullTeamFootballLow}{1.00}
\newcommand{\nullTeamFootballHigh}{1.01}

\newcommand{\teamAfl}{1.12}

\newcommand{\teamAflLowExact}{0.997}
\newcommand{\teamAflHighExact}{1.251}

\newcommand{\teamOtherLow}{0.77}
\newcommand{\teamOtherHigh}{1.12}
\newcommand{\teamGroupsRead}{eight}
\newcommand{\teamGroupsChecked}{nine}

\newcommand{\nflShortWeekNextLow}{10}
\newcommand{\nflShortWeekNextHigh}{11}

\newcommand{\heldLeagueSeasons}{38}

\newcommand{\heldLeaguesIn}{eight}

\newcommand{\heldShort}{1.11}
\newcommand{\heldShortLow}{0.84}
\newcommand{\heldShortHigh}{1.46}
\newcommand{\heldShortNull}{0.96}

\newcommand{\heldGap}{0.93}
\newcommand{\heldGapLow}{0.85}
\newcommand{\heldGapHigh}{1.02}
\newcommand{\heldGapNull}{0.99}
\newcommand{\heldGapNarrow}{0.93}
\newcommand{\heldGapNarrowLow}{0.80}
\newcommand{\heldGapNarrowHigh}{1.07}
\newcommand{\heldGapNarrowNull}{1.00}

\newcommand{\mainWordCount}{2996}
\newcommand{\abstractWordCount}{249}

\newcommand{\narrowcolumns}{\setlength{\tabcolsep}{3pt}}
\newcolumntype{Y}{>{\raggedright\arraybackslash}p{6.0cm}}
\newcolumntype{J}{>{\raggedright\arraybackslash}p{2.0cm}}
\newcolumntype{Z}{>{\raggedright\arraybackslash}p{3.1cm}}

\newcommand{\archivedoi}{10.5281/zenodo.24596753}
\newcommand{\archivelink}{\begingroup\edef\x{\endgroup\noexpand\url{https://doi.org/\archivedoi}}\x}

\begin{document}

\begin{center}
{\Large\bfseries Who is who: comparing injury rates between players and within each
player in professional team sports in Europe, North America, Australia and
Japan\par}
\vspace{1.2em}
{\large Gustavo Pedro Ricou$^{1}$ and Nicholas Mahony$^{2}$\par}
\vspace{0.6em}
$^{1}$School of Computer Science and Statistics, Trinity College Dublin,
Dublin, Ireland\par
$^{2}$Discipline of Anatomy, School of Medicine, Trinity College Dublin,
Dublin, Ireland\par
\end{center}

\vspace{1em}
\noindent\textbf{Corresponding author} Gustavo Pedro Ricou, School of
Computer Science and Statistics, Trinity College Dublin, Dublin, Ireland.
Email: \url{pedrorig@tcd.ie}.

\noindent\textbf{ORCID} Gustavo Pedro Ricou,
\url{https://orcid.org/0009-0001-4196-7213}

\noindent\textbf{Word count} \mainWordCount{} words in the main text;
\abstractWordCount{} in the abstract.

\noindent\textbf{Keywords} Athletic injuries; Sports; Epidemiologic
methods; Selection bias; Population surveillance

\newpage

\begin{abstract}\label{sec:abstract}
\noindent\textbf{Background} Injury rates after heavy weeks, with a full
match or more in the previous seven days, guide player rest and match
calendars. We tested whether comparisons across players and within each
player show the effect of load.

\noindent\textbf{Methods} This observational study used publicly
available match records from \allSports{} professional team sports. Six had
injury records. They were football, Australian rules, basketball, ice
hockey, baseball and American football. Rugby union, rugby league and cricket gave playing time only. In football, injuries were substitutions recorded as due to injury. A simulation replayed the real calendars and
appearances with no effect of load.

\noindent\textbf{Results} Injury records covered \playersDistinct{}
players (\headWomen{} women, in one league) in Europe, North America,
Australia and Japan. In all six sports, comparing players mixed load with
who was picked, because regular starters played most heavy weeks. In the
simulation, comparing each player with their other weeks made heavy weeks
look \nullSeasonAllLow{} to \nullSeasonAllHigh{} times as risky over whole
seasons, because an injury lowers later load. In football, heavy weeks and weeks with no league match had similar rates when both followed a completed match (rate ratio \gapCompletedWithin{}, 95\% confidence interval \gapCompletedWithinLow{} to \gapCompletedWithinHigh{}). Other results are
reported sport by sport.

\noindent\textbf{Conclusions} For prevention, surveillance programmes and
the bodies that set match calendars can check that each comparison within players gives a rate ratio of one in such a simulation.
Club medical staff can read lower injury rates among their most-used
players as a sign of who plays.
\end{abstract}

\section*{What is already known on this topic}\label{sec:box}\label{sec:keypoints}
\begin{itemize}
\item Injury rates that compare players also compare who gets picked,
  because regular starters play most heavy weeks.
\item Comparisons within each player assume that an injury leaves the
  player's later load unchanged.
\end{itemize}

\section*{What this study adds}
\begin{itemize}
\item In every sport with injury records, a simulation with no effect of
  load made each player's heavy weeks look \nullSeasonAllLow{} to \nullSeasonAllHigh{} times as risky as their other weeks over whole seasons.
\end{itemize}

\section*{How might this study affect research, practice, or policy}
\begin{itemize}
\item Researchers, surveillance programmes and the bodies that set match
  calendars can run the same simulation on their own records with the
  program shared with this study.
\item Club medical staff can read lower injury rates among their most-used players as a sign of who plays, rather than of safe load.
\end{itemize}

\section{Background}\label{sec:intro}

Injury rates guide injury prevention in professional team sports.
Surveillance programmes report
them.\cite{ekstrand2021injury_rates_decreased} Leagues and players'
unions use them to plan match
calendars.\cite{fifpro2024_workload,denhollander2024_workload,denhollander2025_calendar_statement}
Club medical staff use them to choose whom to rest. All these uses assume
that a player's recent play changes the risk of injury. We call a player's
play in the previous seven days their recent load. A heavy week has a full
match or more in those seven days.

Injury rates after heavy weeks are compared in two usual ways. The first
compares different players, usually per 1000 hours of
play.\cite{bahr2020ioc,walden2023_football_extension,bengtsson2013congestion,carling2016_congestion,page2023_congestion_review,allen2025_congestion,impellizzeri2023_exposure_dose}
But some players are injured more often than others. Their load and their
previous injuries explain only part of this
difference.\cite{esmaeili2018_individual_risk,lewis2018_nba,hudson2026_concussions}
If the regular starters have most of the heavy weeks, a per-hour rate
compares who gets picked as well as the load.\cite{ricou2026_smf_preprint} In medicine, comparing groups picked in this
way can mislead.\cite{griffith2020_collider}

The second way compares each player with their own other weeks. This
takes out the differences between players. Such a comparison is called a
self-controlled design. Some studies of football have used this design.\cite{carling2016_congestion,hampstead2026_acl} It assumes that an injury leaves the player's later load unchanged.\cite{petersen2016_sccs} But an injury ends the match early and can keep the player out for weeks.

This study measured how large these two biases are in the public match
records of nine professional team sports. A simulation in which load had
no effect on injury checked each comparison. Football, with the fullest
records, gave the worked example.

\section{Methods}

\subsection{Study design and data}\label{sec:design}

This observational study used publicly available match records from nine
professional team sports (Table~1). It follows the Strengthening the
Reporting of Observational Studies in Epidemiology (STROBE)
guideline.\cite{vonelm2007_strobe} The records of men's football come from
the Transfermarkt website, which we call the provider (supplementary
section S2). Supplementary figure S1 shows the study flow.

\subsection{Sports, leagues and players}\label{sec:population}\label{sec:edi}

The nine sports are football, Australian rules football, basketball, ice
hockey, baseball, American football, rugby union, rugby league and
cricket. The first six have injury records. We call one league's seasons
under one substitution limit a league
sample.\cite{ifab2020_five_substitutes} Each analysis uses its own league samples (Table~1). The injury comparison uses the men's leagues of
England, France, Germany, Italy and Spain. The other injury analyses use player groups, such as one league or baseball's pitchers. The women's
injury records come from one league, the Women's National Basketball
Association in the United States.

\subsection{Recent match load}\label{sec:load}

Recent load is a player's league minutes in the
previous seven days, counted in full matches of 90 minutes. A player with
a full match or more had a heavy week. A longer-gap week is a week with no
minutes in which the club's previous league match came \gapShortFrom{} to
\gapShortTo{} days earlier. In such a week the fixture list set the rest,
rather than the coach or an injury. In \heavyWeeklyPct{}\% of heavy weeks, the full match was the normal
weekly match six or seven days earlier.

\subsection{Injury records}\label{sec:injurymethods}

A substitution is injury-coded when the match report gives injury as its
reason (supplementary section S3). We measured the incidence of
injury-coded substitutions per 1000 starter hours, the hours played as a
starter. Within a week, the provider's injury histories showed a recorded
absence after \absenceInjuryPct{}\% of injury-coded substitutions. They
showed one after \absenceTacticalPct{}\% of the other substitutions. We
also counted confirmed injuries. These are injury-coded substitutions followed within a week by such a recorded absence. They are the
nearer of the two counts to the consensus definition of a time-loss
injury.\cite{bahr2020ioc,walden2023_football_extension} In the other
sports, an injury is each league's own record of an absence
(supplementary section S12).

\subsection{Comparisons}\label{sec:comparison}

We compared heavy weeks with longer-gap weeks in two ways. Across
players, we compared different starters' weeks within each league-season
(Mantel--Haenszel method). Within each player, we compared each starter's
heavy weeks with all the same starter's longer-gap weeks in the same
league-season (conditional Poisson regression). We also compared like with
like, using only the longer-gap weeks that followed a completed match. In
a completed match, the starter played all 90 recorded minutes.

A short gap is two to three days since the club's previous match in any
competition. Within each starter, we compared starts after a short gap
with starts six to seven days after the previous match (supplementary
section S6). In all six sports, we also compared each team's games after a short gap with its games after the usual gap. Both kinds of game had the
same number of days to the next game. This comparison per team game counts
every player who played. So it includes rotation (supplementary section
S14).

\subsection{Simulation check}\label{sec:simulation}

A simulation checked each injury comparison (supplementary sections S6 and
S9). It replays the real calendars and each player's real appearances many
times, with no effect of load on injury. Each simulated injury ends its
match early and keeps the player out as long as real absences did. The simulation also builds in team selection and a higher risk after a return from injury. A comparison passes the check when its mean over the runs
lies within two per cent of one and its intervals hold one at least 90\%
of the time. Only comparisons that pass are read as an effect of load.
The others are printed beside their simulated values. We fixed this rule
before the simulation ran.

We built this study in steps, and each step was tested on new data. The
question began in the English Premier League. Its results were then tested
in other European leagues, and then in other sports. In the five leagues
of the injury comparison, we chose the comparison within each starter and
season, confirmed injuries and the longer-gap weeks after a completed
match after earlier runs had given their estimates. So in the last step, the two football comparisons that passed the check were tested in further European leagues whose injury rates by load we had not analysed, under a plan written before the test ran (supplementary section S15).
Further designs, planned in writing from 22 November 2026, are in
supplementary section S14.

\subsection{Statistical analysis}\label{sec:stats}\label{sec:split}\label{sec:keeping}

A statistical model gave each player their own usual injury rate. It
allows for their position, their usual load and how their club records
injuries (supplementary section S11). We split the extra minutes after a
heavy week exactly into a part from who starts and a part from players
playing longer.\cite{kitagawa1955_components} The within-player slope is
the rise in minutes within each starter. We set two criteria on these
minutes before testing them (supplementary section S4). Every other analysis, including the injury comparison, is exploratory.

Most results are incidence rate ratios with their 95\% confidence
intervals (95\% CI). When the interval holds one, we read the two rates as
similar. The analyses used Python (supplementary section S13). The first
author used Anthropic's Claude, a generative artificial intelligence
assistant, to write and check the analysis scripts, search the literature
and edit the text. It was used because it made analyses of this size
possible to build and check. The first author checked every script against hand-worked tests.

\subsection{Patient and public involvement}\label{sec:ppi}

Only the two authors took part in the design, conduct, reporting and
dissemination plans of this research.

\subsection{Ethics}\label{sec:ethics}

The study reused only public,
pre-existing match records. There was no recruitment, no contact with any
individual, no intervention and no access to medical records or clinical
files, so informed consent did not apply. The records name professional
athletes, and some reasons for substitution are injuries. They were
therefore treated as special-category personal data despite being public.
Processing was for scientific research under the research provisions of
the General Data Protection Regulation. Everything published, every table,
figure and deposited file, is aggregated to whole league samples, so none
of it describes a single player. Ethical
approval was not sought. These published records were treated as
archival data, which Trinity College Dublin's research ethics guidance
excepts from approval.\cite{tcd2022_reams_guidance} The same guidance
grants no approval retrospectively.

\section{Results}\label{sec:results}

\subsection{Study sample}\label{sec:sample}

Table~1 gives the data behind each analysis. Injury records covered
\playersDistinct{} players in \shareGroups{} player groups. The injury
comparison in European men's football covered \forcedSeasons{}
league-seasons, with \injuryAppearances{} appearances by \injuryPlayers{}
players.

\begin{table}[htbp]
\centering
\caption{Data for each analysis, by sport and sex, in nine professional team sports}\label{tab:data}
\footnotesize
\narrowcolumns
\begin{tabular}{>{\raggedright\arraybackslash}p{2.2cm}>{\raggedright\arraybackslash}p{3.0cm}rrrrrr>{\raggedright\arraybackslash}p{2.0cm}}
\toprule
Analysis & Sport and sex & \shortstack[r]{League\\samples} & \shortstack[r]{League-\\seasons} & Players & \shortstack[r]{Of them,\\women} & \shortstack[r]{Median\\age, years} & Size & Unit of size \\
\midrule
Players' usual injury rates (Figure S15) & Football, men: England, France, Germany, Italy and Spain, as one group & 1 & 27 & 4591 & 0 & -- & 6710 & injury-coded substitutions, outfield starters \\
 & Australian rules, men & 1 & 9 & 1253 & 0 & -- & 1824 & absences through injury \\
 & Basketball, men and women & 4 & 62 & 4252 & 748 & -- & 3180 & absences through injury; injured-list placements in Japan \\
 & Ice hockey, men & 2 & 13 & 1932 & 0 & -- & 1946 & absences through injury \\
 & Baseball, men, batters and pitchers & 2 & 32 & 3965 & 0 & -- & 4743 & injured-list placements \\
 & American football, men & 1 & 13 & 6580 & 0 & -- & 4613 & absences through injury \\
Who starts (Figure S3A) & Football, 10 men's and 7 women's; rugby union, men & 18 & 74 & 18,419 & 2104 & 24.5--28.7 & 685,148 & appearances \\
Playing time within the same player (Figure S3B) & Football, men and women; rugby union, men; Australian rules, women & 25 & 57 & 18,673 & 3265 & 24.4--28.5 & 14,963 & players with two or more starts \\
The injury comparison (Figure 1 and Table 2) & Football, men, the same five leagues & 5 & 27 & 5942 & 0 & 25.4--27.4 & 296,029 & appearances, goalkeepers included, with 7158 injury-coded substitutions \\
\bottomrule
\end{tabular}

\par\smallskip
\begin{minipage}{\textwidth}\footnotesize
League-seasons, players and sizes are totals over the samples. A player in two samples of one analysis counts twice. The exceptions are two sources
that give each player one identifier. One is the NBA with the G League, its
development league. The other is MLB, for its batters and its pitchers.
The injury comparison counts each player once. One start in
\forcedOneIn{} ended in an injury-coded substitution (supplementary
section S3). The median age column gives the range of the samples' median
ages at appearance (supplementary table S2). A dash marks a row without a
median age. MLB Major League Baseball, NBA National Basketball Association
\end{minipage}
\end{table}

\subsection{Players' usual injury rates}\label{sec:player}

Players differed widely in their usual injury rate, as earlier studies
found.\cite{esmaeili2018_individual_risk,lewis2018_nba} Take two players
picked at random from one season. Over all \shareGroups{} player groups,
the higher of their two usual rates was typically \headAllMedianMrr{}
times the lower (supplementary table S50 and figure S15). Allowing for
each team barely changed this typical ratio (supplementary section S12).

\subsection{Who carried the heavy weeks}\label{sec:whostarts}\label{sec:coaches}

Regular starters had most of the heavy weeks. Who starts accounted for
\compPctLow{}\% to \compPctHigh{}\% of the extra minutes after a heavy week
in all \compPanels{} samples of football and rugby union (supplementary
figure S3A). The first author's nine-sport preprint reported this
share.\cite{ricou2026_arxiv_allocation} The extra minutes were smaller within the same player than across players
in all \otherWithinBelow{} groups of \otherLeagues{} leagues in
\otherSportsRead{} other sports (supplementary section S9).

\subsection{Injury rates after heavy weeks}\label{sec:injuryresults}

In European men's football, the comparison across players made heavy
weeks look \ofGapShortLoadedLowerPct{}\% safer than longer-gap weeks (rate
ratio \ofGapShortLoaded{}, 95\% CI \ofGapShortLoadedLow{} to
\ofGapShortLoadedHigh{}; Figure~1 and Table~2). In the simulation with no effect of load, this
comparison gave \nullAcrossLow{} to \nullAcrossHigh{}. It gave
\nullAcrossSelection{} when players at lower risk carried more load
(supplementary section S6).

Comparisons within each player and season leaned the other way. An injury
lowers later load. So in the simulation, each player's heavy weeks looked riskier than their other weeks. Over whole seasons they
looked \nullWithinFootballLow{} to \nullWithinFootballHigh{} times as risky
in football, and \nullSeasonOtherLow{} to \nullSeasonOtherHigh{} times as
risky in the other sports (supplementary sections S6 and S9). Only football's like-with-like comparisons gave about one in the simulation.

Of heavy weeks, \heavyCompletedPct{}\% followed a match the starter
completed, against \longGapCompletedPct{}\% of longer-gap weeks. When both
kinds of week
followed a completed match, their rates were similar on both counts
(Figure~1 and Table~2). On injury-coded substitutions, the rate ratio within each starter and season was \gapCompletedWithin{} (95\% CI \gapCompletedWithinLow{} to
\gapCompletedWithinHigh{}). The rates stayed similar with every club match
counted, and without each club's first league match after a break such as
an international break. One version across starters made heavy weeks look riskier (supplementary section S6).

\begin{table}[htbp]
\centering
\caption{Injury comparisons for heavy weeks and short gaps in European men's football, in \forcedSeasons{} league-seasons of England, France, Germany, Italy and Spain, and in \heldLeagueSeasons{} of \heldLeaguesIn{} other European leagues}\label{tab:comparisons}
\footnotesize
\begin{tabular}{YJZZ}
\toprule
Comparison & Outcome & Rate ratio (95\% CI) & With no effect of load \\
\midrule
Across players, every longer-gap week & Injury-coded & \ofGapShortLoaded{} (\ofGapShortLoadedLow{} to \ofGapShortLoadedHigh{}) & \nullAcrossLow{} to \nullAcrossHigh{} \\
 & Confirmed & \ofGapShortNarrow{} (\ofGapShortNarrowLow{} to \ofGapShortNarrowHigh{}) & \\
Within each player and season, every longer-gap week & Injury-coded & \gapWithin{} (\gapWithinLow{} to \gapWithinHigh{}) & \nullAdvisedLow{} to \nullAdvisedHigh{} \\
 & Confirmed & \gapWithinNarrow{} (\gapWithinNarrowLow{} to \gapWithinNarrowHigh{}) & \\
Across players, longer-gap weeks after a completed match & Injury-coded & \gapCompletedAcross{} (\gapCompletedAcrossLow{} to \gapCompletedAcrossHigh{}) & \nullCompletedAcrossLow{} to \nullCompletedAcrossHigh{} \\
 & Confirmed & \gapCompletedAcrossNarrow{} (\gapCompletedAcrossNarrowLow{} to \gapCompletedAcrossNarrowHigh{}) & \\
Within each player and season, longer-gap weeks after a completed match & Injury-coded & \gapCompletedWithin{} (\gapCompletedWithinLow{} to \gapCompletedWithinHigh{}) & \nullCompletedLow{} to \nullCompletedHigh{} \\
 & Confirmed & \gapCompletedWithinNarrow{} (\gapCompletedWithinNarrowLow{} to \gapCompletedWithinNarrowHigh{}) & \\
The same, in \heldLeaguesIn{} other European leagues & Injury-coded & \heldGap{} (\heldGapLow{} to \heldGapHigh{}) & \heldGapNull{} to \heldGapNarrowNull{} \\
 & Confirmed & \heldGapNarrow{} (\heldGapNarrowLow{} to \heldGapNarrowHigh{}) & \\
Within each player and season, a short gap against six to seven days & Confirmed & \gapTwoThreeNarrow{} (\gapTwoThreeNarrowLow{} to \gapTwoThreeNarrowHigh{}) & \nullTwoThreeNarrowLow{} \\
Within each player and season, a short gap after a completed match & Confirmed & \shortCompletedNarrow{} (\shortCompletedNarrowLow{} to \shortCompletedNarrowHigh{}) & \nullShortCompletedNarrowLow{} to \nullShortCompletedNarrowHigh{} \\
The same, in \heldLeaguesIn{} other European leagues & Confirmed & \heldShort{} (\heldShortLow{} to \heldShortHigh{}) & \heldShortNull{} \\
Per team game, a short gap against the usual gap & Confirmed & \teamFootball{} (\teamFootballLow{} to \teamFootballHigh{}) & \nullTeamFootballLow{} to \nullTeamFootballHigh{} \\
\bottomrule
\end{tabular}
\par\smallskip
\begin{minipage}{\textwidth}\footnotesize
Injury-coded counts injury-coded substitutions, and confirmed counts those
followed by a recorded absence. A heavy week had a full league match or
more in the previous seven days. A longer-gap week followed a gap of
\gapShortFrom{} to \gapShortTo{} days between league matches. A short gap
was two to three days since the club's previous match in any competition.
The last column gives the range of the comparison's mean over the settings
of the simulation with no effect of load (supplementary sections S6 and
S14). For a comparison with two rows, that range covers both outcomes. The
comparison within each player against every longer-gap week, and the
short gap in the other European leagues, are printed beside these values,
as the Methods ask of a comparison that misses the check (supplementary
section S15). CI confidence interval
\end{minipage}
\end{table}

\begin{figure}[htbp]
\centering
\includegraphics{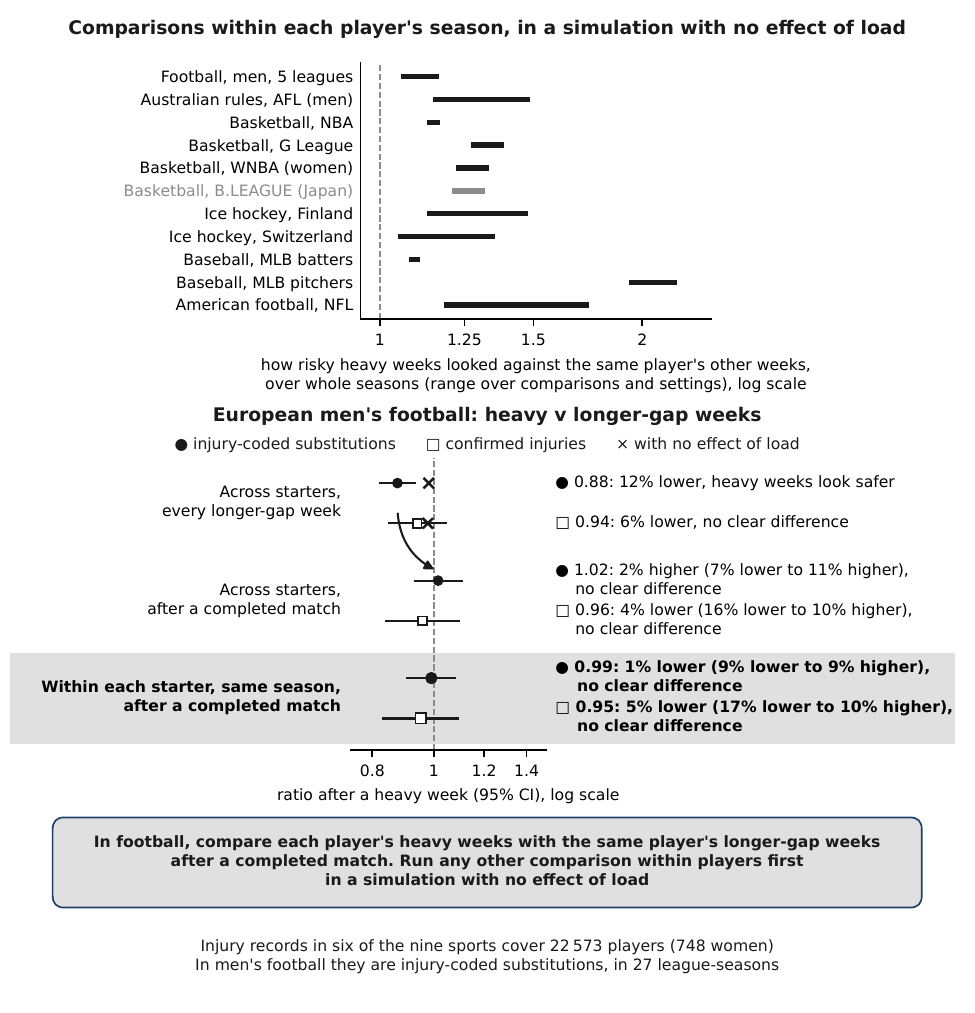}
\label{fig:central}\label{fig:injury}\caption{Heavy weeks compared within each player in a simulation with no effect of load, in \shareGroups{} player groups of \shareSports{} sports, and heavy weeks against longer-gap weeks in European men's football, in \forcedSeasons{} league-seasons. The top row shows how comparisons within each player read when load has no effect. For each player group, the line spans how risky each player's heavy weeks looked against their other weeks over whole seasons, across the comparisons and settings of the simulation (supplementary tables S22 and S45). A fair comparison would give one, the dashed line. Japan's B.LEAGUE is drawn lighter because it had only \bleagueInjuries{} injuries. The lower row shows how to compare like with like. A filled mark shows the rate ratio of outfield starters' injury-coded substitutions per starter hour. An open mark shows the rate ratio of their confirmed injuries. Goalkeepers' \ofKeeperStarts{} starts are left out. The first pair is the comparison across starters against every longer-gap week. A cross marks its value in the simulation. The other pairs use only longer-gap weeks after a completed match, across starters and within each starter (shaded) in the same league-season. The words beside each interval give its reading. They say safer or riskier when the interval lies wholly below or above one, and no clear difference when it holds one. The arrow runs from the comparison across starters to the comparisons after a completed match. A heavy week had a full league match or more in the previous seven days. A longer-gap week followed a gap of \gapShortFrom{} to \gapShortTo{} days between the club's league matches. The box gives this study's own advice. AFL Australian Football League, CI confidence interval, MLB Major League Baseball, NBA National Basketball Association, NFL National Football League, v versus, WNBA Women's National Basketball Association}
\end{figure}

\subsection{Short gaps}\label{sec:shortgaps}

Within each starter and season, starts two to three days after the
previous match had more confirmed injuries than starts six to seven days
after it. The rate ratio was \gapTwoThreeNarrow{} (95\% CI
\gapTwoThreeNarrowLow{} to \gapTwoThreeNarrowHigh{}).
In the simulation with no effect of load, this ratio was
\nullTwoThreeNarrowLow{}. When the starter had completed that match, the
ratio was \shortCompletedNarrow{} (95\% CI \shortCompletedNarrowLow{} to
\shortCompletedNarrowHigh{}). Here the simulation gave about one. On
injury-coded substitutions both ratios were smaller (supplementary section
S6). Supplementary section S14 shows how strong a hidden cause would need
to be to explain them away.

Per team game, football's league matches after two to three days had \teamFootball{} times the confirmed injuries of those after six to seven days (95\% CI \teamFootballLow{} to \teamFootballHigh{}). In the \teamGroupsRead{} other
groups with an estimate, the ratios ran from \teamOtherLow{} to
\teamOtherHigh{}. The highest was the Australian Football League's
\teamAfl{} (95\% CI \teamAflLowExact{} to \teamAflHighExact{}). In the
simulation with no effect of load, this comparison gave about one in all
\teamGroupsChecked{} groups.

In \heldLeaguesIn{} other European leagues, tested afterwards, heavy weeks
and longer-gap weeks after a completed match again had similar rates. The
rate ratio was \heldGap{} on injury-coded substitutions (95\% CI
\heldGapLow{} to \heldGapHigh{}). The short-gap comparison missed the
simulation check there, so it gives no reading (Table~2 and supplementary
section S15).

\subsection{Results by sport}\label{sec:bysport}

Table~3 gives the results sport by sport.

\begin{table}[htbp]
\centering
\caption{Results by sport in the \shareGroups{} player groups with injury records}\label{tab:bysport}
\footnotesize
\narrowcolumns
\begin{tabular}{>{\raggedright\arraybackslash}p{2.5cm}>{\raggedright\arraybackslash}p{2.3cm}rr>{\raggedright\arraybackslash}p{2.3cm}>{\raggedright\arraybackslash}p{2.8cm}}
\toprule
League & What counts as an injury & \shortstack[r]{Two players:\\median rate\\ratio (95\% CI)} & \shortstack[r]{Within each\\player, with no\\effect of load} & Heavy weeks within each player, like with like (95\% CI) & Short gap, per team game (95\% CI) \\
\midrule
\multicolumn{6}{l}{\textit{Football}} \\
European men's football & Injury-coded substitution & 1.83 (1.76 to 1.90) & 1.06 to 1.17 & 0.99 (0.91 to 1.09), no clear difference & 1.19 (1.03 to 1.36), riskier \\
\multicolumn{6}{l}{\textit{Basketball}} \\
B.LEAGUE, Japan & Injured-list placement & 2.84 (1.00 to 6.90) & 1.21 to 1.32 & none passed the check & too few injuries \\
G League & Missed next game, box score & 2.04 (1.78 to 2.32) & 1.27 to 1.39 & none passed the check & 0.88 (0.74 to 1.06), no clear difference \\
NBA & Missed next game, box score & 1.97 (1.83 to 2.14) & 1.13 to 1.17 & none passed the check & 0.94 (0.85 to 1.05), no clear difference \\
WNBA (women) & Missed next game, box score & 1.82 (1.61 to 2.08) & 1.22 to 1.33 & none passed the check & 0.95 (0.70 to 1.27), no clear difference \\
\multicolumn{6}{l}{\textit{Baseball}} \\
MLB, batters & Injured-list placement & 1.71 (1.62 to 1.81) & 1.08 to 1.11 & none passed the check & 1.03 (0.92 to 1.15), no clear difference \\
MLB, pitchers & Injured-list placement & 1.59 (1.47 to 1.72) & 1.93 to 2.19 & none passed the check & 1.03 (0.93 to 1.15), no clear difference \\
\multicolumn{6}{l}{\textit{Australian rules}} \\
AFL, Australia (men) & Missed next game, team list & 1.66 (1.54 to 1.79) & 1.15 to 1.49 & none passed the check & 1.12 (0.997 to 1.251), no clear difference \\
\multicolumn{6}{l}{\textit{Ice hockey}} \\
Liiga, Finland & Missed next game, club roster & 4.16 (3.11 to 5.84) & 1.13 to 1.48 & none passed the check & 0.77 (0.47 to 1.24), no clear difference \\
National League, Switzerland & Missed next game, game sheet & 1.79 (1.66 to 1.93) & 1.05 to 1.35 & none passed the check & 0.99 (0.85 to 1.14), no clear difference \\
\multicolumn{6}{l}{\textit{American football}} \\
NFL & Missed next game, club injury report & 1.90 (1.81 to 2.00) & 1.18 to 1.74 & none passed the check & missed the check \\
\bottomrule
\end{tabular}

\par\smallskip
\begin{minipage}{\textwidth}\footnotesize
Each row is a league or player group with an injury record. Two players
gives the median rate ratio between two players picked at random from one
season, the higher usual injury rate over the lower (supplementary table
S50). Within each player comes from the simulation with no effect of load.
It shows how risky each player's heavy weeks looked against their other
weeks over whole seasons. It is the range over comparisons and settings
(supplementary tables S22 and S45). A comparison passes the check when it
gives about one in that simulation. Like with like sets each starter's
heavy weeks against the same starter's longer-gap weeks when both followed
a completed match (Table~2). Only football had a comparison within players
that passed the check (supplementary section S9). Per team game sets each
team's games after a short gap against the same team's games after its
usual gap, in the same season (supplementary table S59). In football the
short gap is two to three days, against six to seven. In basketball and
ice hockey it is the second game of a back-to-back, against games after
one or two rest days. In Australian rules it is a break of six days or
less, against seven to eight. In baseball it is a game the day after the
last one, against one after an off day. In American football it is five
days or less, against six to eight. Football's row counts injury-coded
substitutions. Its range within each player covers confirmed injuries too.
Per team game, it counts confirmed injuries only. The words give each interval's reading. They say riskier when the interval lies wholly above one, and no clear difference when it holds one. Japan's B.LEAGUE had few injuries. So its
median rate ratio has a wide interval. Finland's Liiga has the highest
median rate ratio because its teams record injuries in different ways
(supplementary section S12). AFL Australian Football League, CI confidence
interval, MLB Major League Baseball, NBA National Basketball Association,
NFL National Football League, WNBA Women's National Basketball Association
\end{minipage}
\end{table}

\section{Discussion}\label{sec:discussion}

Both usual comparisons mixed load with something else. Comparing players
mixed it with who gets picked. This made football's heavy weeks look safer than longer-gap weeks.
Comparing each player with their own weeks mixed it with what an injury
does to later play. So with no effect of load, heavy weeks looked riskier than other weeks in every sport with injury records. Compared like with like, football's heavy and longer-gap weeks had similar rates.

\subsection{Comparing players}\label{sec:dividing}\label{sec:cost}

Team selection makes heavy weeks look safe. Our earlier preprint argued that a per-hour rate compares players as well as loads.\cite{ricou2026_smf_preprint} Most heavy weeks belong to regular starters. Workplace health research calls this the healthy worker effect. Those who do most of a job are those fit enough to stay in it.\cite{li1999_healthy_worker,steenland1996_survivor}

So a healthy worker effect credits heavy weeks with the regular starters'
lower risk. Regular match play could also protect players over months.

\subsection{Comparing each player with their own weeks}\label{sec:research}

An injury makes heavy weeks look risky within each player. Self-controlled
designs assume that later load stays the same after an
injury.\cite{petersen2016_sccs} This assumption can be
tested.\cite{whitaker2018_sccs_assumptions} One method still works when the assumption fails, but only for brief exposures and rare events that happen
once.\cite{farrington2009_sccs_perturbed} In sport, injuries happen
again, and load comes every week.

So surveillance programmes and researchers can run each comparison within
players in a simulation with no effect of load before they read it.
Outside football, no reference weeks that we tried passed
(supplementary section S9). A comparison per team game passed in every sport but American football. Several other designs failed (supplementary section S14). The program shared with this study runs the
check on a club's or a league's own records. Studies of public match
records can also count confirmed injuries and every match played.

\subsection{Implications for prevention and policy}\label{sec:clinical}\label{sec:elements}

Leagues and players' unions can weigh short gaps after a completed match
when they plan match calendars. In football, starts two to three days
after a completed match had about \shortCompletedExcessPerThousand{} more
confirmed injuries per 1000 league starts than those after six to seven days (\shortCompletedExcessPerThousandLow{} to \shortCompletedExcessPerThousandHigh{}). At an average club, that is one
more such injury every \shortCompletedExcessClubSeasons{} seasons. This count sees only injuries that ended a player's match and were then
listed as absences. So a club's whole excess may be larger. In the other sports, short and
usual gaps had similar rates per team game. That shows present practice,
not what less rest would do (supplementary section S14).

For club medical staff, a lower per-hour injury rate among their most-used
players can show who plays. It need not mean that their load is safe. Compared like with like, heavy and longer-gap weeks had similar rates (rate ratio
\gapCompletedWithin{}, 95\% CI \gapCompletedWithinLow{} to
\gapCompletedWithinHigh{}). So clubs can aim rest and rotation at short
gaps after a completed match.

A decision to rest a player can also weigh five other things. They are
previous injury,\cite{hagglund2006_previous_injury} the recovery time
between matches,\cite{bengtsson2013congestion,page2023_congestion_review}
every match and the travel between them,\cite{allen2025_congestion}
training load,\cite{impellizzeri2023_exposure_dose} and squad depth and
rotation.\cite{carling2016_congestion,carling2015_congestion_exposure}
Match records lack others, such as a knock (a minor injury that a player plays through) and the reason a player was rested.

\subsection{Limitations}\label{sec:limitations}\label{sec:strengths}

The most important limitation is how recent load was measured.\cite{wang2024_misclassification} The records hold no training and no national-team matches in the international breaks. So an international player's
load is
understated in exactly the weeks that congestion research studies
(supplementary table S12). Leaving out the league matches just after these
breaks kept the rates within each starter similar (supplementary section
S6). In football and rugby union, recorded minutes also stop at the full
match. This sets the smallest within-player slope the design could detect
(supplementary section S6).

Second, the records are public match data, not surveillance data kept by
club medical staff.\cite{hoenig2022_citizen_science,krutsch2020_media}
Club medical records are private, so no injury could be checked against
them. The count sees only injuries that end a player's match. It misses
injuries in training and injuries that players play through. Outside
football, some lists record rest or a move to another team as an injury.
No video audit of the study seasons was feasible. But injury-coded
substitutions were confirmed about as often after heavy weeks as after
longer-gap weeks, at \absLoadedStarterPct{}\% against \absLongGapPct{}\%
(supplementary section S6).

Third, the design is observational.\cite{shrier2022_causal_workload} The
simulation holds only the causes it builds in. A player may carry a knock
through a completed match, and staff may rest players at rising
risk.\cite{krishna2026_workload} The strongest knock we simulated could
hide a rise of up to about \knockMaxPct{}\% (supplementary section S6).

Fourth, we saved the two pre-specified criteria, with a date, to the
project's own files, not to a public protocol. We saved both after our
earlier preprint had used the same records. The first passed narrowly and
the second failed (supplementary section S4). In the other European
leagues, the heavy-week test could often miss a rise of one in ten
(supplementary section S15). We have registered the two football
comparisons that passed the simulation check as predictions for the next
seasons, for anyone to test when those seasons end.\cite{ricou2026_osf_prediction}

Finally, the samples differ in sport, sex, era and size. The comparison
within players
comes from European men's football (supplementary section S9). Per team
game, the National Football League had too few games to compare, because
games on a short week are mostly followed by \nflShortWeekNextLow{} or
\nflShortWeekNextHigh{} days. The B.LEAGUE has too few injuries
(supplementary section S14). Sports also differ in how their records count
injuries and in how teams rest players. These differences may explain why
some results depend on the sport (Table~3).

\section{Conclusions}\label{sec:conclusion}

Rates that compare players also compare who gets picked.
Comparisons within each player are biased, because an injury lowers later
load. For prevention, surveillance programmes and the bodies that set
match calendars can first run each comparison within players in a
simulation with no effect of load. In football, compared like with like, heavy and longer-gap weeks had similar rates. Per team game, short gaps had more confirmed injuries than usual gaps in football and similar rates in the other sports. Club
medical staff can read lower injury rates among their most-used players as
a sign of who plays.

\section*{Declarations}\label{sec:declarations}

\noindent\textbf{Contributors} GPR designed the study, collected and
analysed the data and wrote the manuscript. NM shaped the methods and the
clinical framing of this manuscript. GPR used a generative artificial
intelligence assistant, as the Methods describe. Both authors read and
approved the final version. GPR is the guarantor.

\noindent\textbf{Funding} This research received no specific grant from
any funding agency in the public, commercial or not-for-profit sectors.

\noindent\textbf{Competing interests} None declared.

\noindent\textbf{Patient consent for publication} Not applicable.

\noindent\textbf{Data availability statement} Data are available in a
public, open access repository. The aggregated results behind every table
and figure, and the analysis scripts, are in the archived record at
\archivelink{}. The record also holds tools/paper\_a\_bias\_check.py. This
program runs the simulation check on other records. No player-level
record is shared. The scripts read match records that identify players.
So these records are not redistributed. The archive documents how to
rebuild these records from the public providers. It also gives each
source's terms of use as we read them and any permission we hold.
Statistics from NBA.com are credited to NBA.com, as its terms ask.

\noindent\textbf{Prior dissemination} A version of this manuscript is
posted on arXiv as version 6 of our earlier preprint, arXiv:2608.11228,
with the DOI \url{https://doi.org/10.48550/arXiv.2608.11228}. It is posted
under arXiv's non-exclusive licence to distribute, version 1.0. That
preprint raised the question this manuscript answers. Some of this
manuscript's results were first posted in the first author's nine-sport
preprint, arXiv:2609.05584v1.

\bibliographystyle{bmj_vancouver}
\bibliography{references}

\begin{thebibliography}{10}

\bibitem{ekstrand2021injury_rates_decreased}
Ekstrand J, Spreco A, Bengtsson H, et~al.
\newblock Injury rates decreased in men's professional football: an 18-year
  prospective cohort study of almost 12 000 injuries sustained during 1.8
  million hours of play.
\newblock Br J Sports Med 2021;55:1084--91.
\newblock https://doi.org/10.1136/bjsports-2020-103159.

\bibitem{fifpro2024_workload}
{FIFPRO}. Excessive workload demands: player performance, recovery and health.
  {FIFPRO} {Player} {Workload} {Monitoring} men's football report, 2023/24
  season; 2024.
\newblock With Football Benchmark; the core group of players, page 15. Accessed
  28 September 2026.
\newblock Available from:
  \url{https://media.fifpro.org/media/1oxkdnt2/en_pwm_men-s-annual-report-2024.pdf}.

\bibitem{denhollander2024_workload}
den Hollander S, Kerkhoffs G, Gouttebarge V.
\newblock The impact of match workload and international travel on injuries in
  professional men's football.
\newblock Sports (Basel) 2024;12:212.
\newblock https://doi.org/10.3390/sports12080212.

\bibitem{denhollander2025_calendar_statement}
den Hollander S, Burgess D, Kerkhoffs G, et~al.
\newblock The {International} {Match} {Calendar} in men's professional
  football: an expert position statement.
\newblock Scand J Med Sci Sports 2025;35:e70163.
\newblock https://doi.org/10.1111/sms.70163.

\bibitem{bahr2020ioc}
Bahr R, Clarsen B, Derman W, et~al.
\newblock {International Olympic Committee} consensus statement: methods for
  recording and reporting of epidemiological data on injury and illness in
  sport 2020 (including {STROBE} Extension for Sport Injury and Illness
  Surveillance ({STROBE-SIIS})).
\newblock Br J Sports Med 2020;54:372--89.
\newblock https://doi.org/10.1136/bjsports-2019-101969.

\bibitem{walden2023_football_extension}
Wald{\'e}n M, Mountjoy M, McCall A, et~al.
\newblock Football-specific extension of the {IOC} consensus statement: methods
  for recording and reporting of epidemiological data on injury and illness in
  sport 2020.
\newblock Br J Sports Med 2023;57:1341--50.
\newblock https://doi.org/10.1136/bjsports-2022-106405.

\bibitem{bengtsson2013congestion}
Bengtsson H, Ekstrand J, H{\"a}gglund M.
\newblock Muscle injury rates in professional football increase with fixture
  congestion: an 11-year follow-up of the {UEFA} {Champions} {League} injury
  study.
\newblock Br J Sports Med 2013;47:743--7.
\newblock https://doi.org/10.1136/bjsports-2013-092383.

\bibitem{carling2016_congestion}
Carling C, McCall A, Le~Gall F, et~al.
\newblock The impact of short periods of match congestion on injury risk and
  patterns in an elite football club.
\newblock Br J Sports Med 2016;50:764--8.
\newblock https://doi.org/10.1136/bjsports-2015-095501.

\bibitem{page2023_congestion_review}
Page RM, Field A, Langley B, et~al.
\newblock The effects of fixture congestion on injury in professional male
  soccer: a systematic review.
\newblock Sports Med 2023;53:667--85.
\newblock https://doi.org/10.1007/s40279-022-01799-5.

\bibitem{allen2025_congestion}
Allen T, Taberner M, Zhilkin M, et~al.
\newblock Evolving running load demands and fixture congestion in the {English}
  {Premier} {League}: a decade of insights from 2015/2016 to 2024/2025.
\newblock Br J Sports Med 2025;59:1643--54.
\newblock https://doi.org/10.1136/bjsports-2025-110430.

\bibitem{impellizzeri2023_exposure_dose}
Impellizzeri FM, Shrier I, McLaren SJ, et~al.
\newblock Understanding training load as exposure and dose.
\newblock Sports Med 2023;53:1667--79.
\newblock https://doi.org/10.1007/s40279-023-01833-0.

\bibitem{esmaeili2018_individual_risk}
Esmaeili A, Hopkins WG, Stewart AM, et~al.
\newblock The individual and combined effects of multiple factors on the risk
  of soft tissue non-contact injuries in elite team sport athletes.
\newblock Front Physiol 2018;9:1280.
\newblock https://doi.org/10.3389/fphys.2018.01280.

\bibitem{lewis2018_nba}
Lewis M.
\newblock It's a hard-knock life: game load, fatigue, and injury risk in the
  {National Basketball Association}.
\newblock J Athl Train 2018;53:503--9.
\newblock https://doi.org/10.4085/1062-6050-243-17.

\bibitem{hudson2026_concussions}
Hudson SJ, Williams S, Starling L, et~al.
\newblock Training-related concussions in men's {English} {Premiership} {Rugby}
  {Union}: an eight-season prospective cohort study of 153 243 player-weeks.
\newblock Br J Sports Med 2026;60:1394--400.
\newblock https://doi.org/10.1136/bjsports-2026-111653.

\bibitem{ricou2026_smf_preprint}
Pedro~Ricou G, Mahony N. Playing time tracks recent exposure in fifteen men's
  and women's football leagues: the denominator gradient and what it does to
  per-minute injury rates; 2026.
\newblock {arXiv}:2608.11228v3 [stat.AP]. Not peer reviewed.
\newblock Available from: \url{https://arxiv.org/abs/2608.11228v3}.
  https://doi.org/10.48550/arXiv.2608.11228.

\bibitem{griffith2020_collider}
Griffith GJ, Morris TT, Tudball MJ, et~al.
\newblock Collider bias undermines our understanding of {COVID-19} disease risk
  and severity.
\newblock Nat Commun 2020;11:5749.
\newblock https://doi.org/10.1038/s41467-020-19478-2.

\bibitem{hampstead2026_acl}
Hampstead W, Pappas E, Whalan M, et~al.
\newblock Match exposure and estimated travel before {ACL} injury in
  professional women's football: an exploratory within-player case-crossover
  study.
\newblock Sports Med Open 2026;12:170.
\newblock https://doi.org/10.1186/s40798-026-01144-4.

\bibitem{petersen2016_sccs}
Petersen I, Douglas I, Whitaker H.
\newblock Self controlled case series methods: an alternative to standard
  epidemiological study designs.
\newblock BMJ 2016;354:i4515.
\newblock https://doi.org/10.1136/bmj.i4515.

\bibitem{vonelm2007_strobe}
von Elm E, Altman DG, Egger M, et~al.
\newblock The {Strengthening} the {Reporting} of {Observational} {Studies} in
  {Epidemiology} ({STROBE}) statement: guidelines for reporting observational
  studies.
\newblock Lancet 2007;370:1453--7.
\newblock https://doi.org/10.1016/S0140-6736(07)61602-X.

\bibitem{ifab2020_five_substitutes}
{The IFAB}. Five substitutes option temporarily allowed for competition
  organisers (8 May 2020); and The {IFAB} permanently approves five-substitute
  option in top-level competitions (13 June 2022); 2022.
\newblock The 2020 announcement:
  \url{https://www.theifab.com/news/five-substitutes-option-temporarily-allowed-for-competition-organisers/}.
  Both accessed 27 September 2026.
\newblock Available from:
  \url{https://www.theifab.com/news/the-ifab-permanently-approves-five-substitute-option-in-top-level-competitions/}.

\bibitem{kitagawa1955_components}
Kitagawa EM.
\newblock Components of a difference between two rates.
\newblock J Am Stat Assoc 1955;50:1168--94.
\newblock https://doi.org/10.2307/2281213.

\bibitem{tcd2022_reams_guidance}
{Trinity College Dublin}. Guidance for using the {Research} {Ethics}
  {Application} {Management} {System} ({REAMS}): making an application; 2022.
\newblock Institutional guidance, pages 5 and 27--28. Accessed 25 September
  2026.
\newblock Available from:
  \url{https://www.tcd.ie/research/support/assets/pdf/Top%20Guidance_Simplified-31st%20Oct.pdf}.

\bibitem{ricou2026_arxiv_allocation}
Pedro~Ricou G. One equation allocates playing time in competitive team sports:
  the rulebook sets the arithmetic and the league sets the habit; 2026.
\newblock {arXiv}:2609.05584v1 [physics.soc-ph]. Not peer reviewed.
\newblock Available from: \url{https://arxiv.org/abs/2609.05584v1}.
  https://doi.org/10.48550/arXiv.2609.05584.

\bibitem{li1999_healthy_worker}
Li CY, Sung FC.
\newblock A review of the healthy worker effect in occupational epidemiology.
\newblock Occup Med (Lond) 1999;49:225--9.
\newblock https://doi.org/10.1093/occmed/49.4.225.

\bibitem{steenland1996_survivor}
Steenland K, Deddens J, Salvan A, et~al.
\newblock Negative bias in exposure-response trends in occupational studies:
  modeling the healthy worker survivor effect.
\newblock Am J Epidemiol 1996;143:202--10.
\newblock https://doi.org/10.1093/oxfordjournals.aje.a008730.

\bibitem{whitaker2018_sccs_assumptions}
Whitaker HJ, Ghebremichael-Weldeselassie Y, Douglas IJ, et~al.
\newblock Investigating the assumptions of the self-controlled case series
  method.
\newblock Stat Med 2018;37:643--58.
\newblock https://doi.org/10.1002/sim.7536.

\bibitem{farrington2009_sccs_perturbed}
Farrington CP, Whitaker HJ, Hocine MN.
\newblock Case series analysis for censored, perturbed, or curtailed post-event
  exposures.
\newblock Biostatistics 2009;10:3--16.
\newblock https://doi.org/10.1093/biostatistics/kxn013.

\bibitem{hagglund2006_previous_injury}
H{\"a}gglund M, Wald{\'e}n M, Ekstrand J.
\newblock Previous injury as a risk factor for injury in elite football: a
  prospective study over two consecutive seasons.
\newblock Br J Sports Med 2006;40:767--72.
\newblock https://doi.org/10.1136/bjsm.2006.026609.

\bibitem{carling2015_congestion_exposure}
Carling C, McCall A, Le~Gall F, et~al.
\newblock What is the extent of exposure to periods of match congestion in
  professional soccer players?
\newblock J Sports Sci 2015;33:2116--24.
\newblock https://doi.org/10.1080/02640414.2015.1091492.

\bibitem{wang2024_misclassification}
Wang C, Kaufman JS, Shrier I.
\newblock Exposure misclassification: an ``immortal time'' bias in
  observational studies of training load and injury.
\newblock J Clin Epidemiol 2024;176:111581.
\newblock https://doi.org/10.1016/j.jclinepi.2024.111581.

\bibitem{hoenig2022_citizen_science}
Hoenig T, Edouard P, Krause M, et~al.
\newblock Analysis of more than 20,000 injuries in {European} professional
  football by using a citizen science-based approach: an opportunity for
  epidemiological research?
\newblock J Sci Med Sport 2022;25:300--5.
\newblock https://doi.org/10.1016/j.jsams.2021.11.038.

\bibitem{krutsch2020_media}
Krutsch V, Grechenig S, Loose O, et~al.
\newblock Injury analysis in professional soccer by means of media reports:
  only severe injury types show high validity.
\newblock Open Access J Sports Med 2020;11:123--31.
\newblock https://doi.org/10.2147/OAJSM.S251081.

\bibitem{shrier2022_causal_workload}
Shrier I, Wang C, Stokes T, et~al.
\newblock Causal effects, workload and injury risk: the importance of
  specifying the research question.
\newblock J Sci Med Sport 2022;25:574--8.
\newblock https://doi.org/10.1016/j.jsams.2022.03.018.

\bibitem{krishna2026_workload}
Krishna K, Lychagin S, McLean LS.
\newblock Injury, workload, and the strategic response of coaches in the {NBA}.
\newblock Cambridge, MA: National Bureau of Economic Research; 2026. 35803.
\newblock https://doi.org/10.3386/w35803.

\bibitem{ricou2026_osf_prediction}
Pedro~Ricou G. Two injury comparisons in European football: a prediction for
  the 2026-27 seasons; 2026.
\newblock Registered 28 November 2026, under embargo until 1 July 2027.
\newblock OSF Registries.
\newblock Available from: \url{https://osf.io/y8zkg}.

\end{thebibliography}

\end{document}